\documentclass[aps,prd,secnumarabic,amssymb, amsmath,nobibnotes,nofootinbib,11pt]{revtex4} 
\usepackage{amsfonts,amsmath,hyperref,url, color}
\usepackage{bm, bbm}
\usepackage{graphicx}
\usepackage{mathtools}

\newcommand{\be}{\begin{equation}}
\newcommand{\bal}{\begin{align}}
\newcommand{\eal}{\end{align}}
\newcommand{\ee}{\end{equation}}
\newcommand{\bea}{\begin{eqnarray}}
\newcommand{\eea}{\end{eqnarray}}
\newcommand{\bit}{\begin{itemize}}
\newcommand{\eit}{\end{itemize}}

\newcommand{\braket}[2]{\left\langle#1 |  #2\right\rangle}

\newcommand{\ket}[1]{\left|#1\right\rangle}

\usepackage{color}

\begin{document}

\title{Conformal quantum mechanics of causal diamonds}

\author{Michele Arzano}
\email{michele.arzano@na.infn.it}
\affiliation{Dipartimento di Fisica ``E. Pancini", Universit\`a di Napoli Federico II, I-80125 Napoli, Italy\\}
\affiliation{INFN, Sezione di Napoli,\\ Complesso Universitario di Monte S. Angelo,\\
Via Cintia Edificio 6, 80126 Napoli, Italy}

\begin{abstract}
It is shown that a general radial conformal Killing vector in Minkowski space-time can be associated to a generator of time evolution in conformal quantum mechanics. Among these conformal Killing vectors one finds a class which maps causal diamonds in Minkowski space-time into themselves. The flow of such Killing vectors describes worldlines of accelerated observers with a finite lifetime within the causal diamond. Time evolution of static diamond observers is equivalent to time evolution in conformal quantum mechanics governed by a hyperbolic Hamiltonian and covering only a segment of the time axis. This indicates that the Unruh temperature perceived by static diamond observers in the vacuum state of inertial observers in Minkowski space can be obtained from the behaviour of the two-point functions of conformal quantum mechanics.  
\end{abstract}

\maketitle

\section{Introduction}
Do static observers with a finite lifetime in Minkowski space perceive the quantum vacuum as a thermal state? In this note we unveil a connection between the symmetries of {\it causal diamonds}, regions in Minkowski space-time defined by the intersection of past and future light cones of two events, and time evolution in conformal quantum mechanics, a $0+1$-dimensional field theory with conformal invariance \cite{deAlfaro:1976vlx}. This connection provides group theoretic evidence for a {\it diamond temperature} associated to static observers within the diamond. 

Our results are based on a simple observation: the most general radial conformal Killing vector in Minkowski space-time, in space-time dimensions $d\geq 3$, is equivalent, at a group theoretic level, to the most general Hamiltonian governing time evolution in conformal quantum mechanics. A special class of {\it hyperbolic} radial conformal Killing vectors generates time evolution within the causal diamond which correspond to a choice of Hamiltonian in conformal quantum mechanics alternative to ordinary time translations. These translations cover only a finite region on the time domain of the theory whose size matches the size of the causal diamond in Minkowski space. The existence of different notions of time translations on both sides, leads to two-point functions which are thermal for observers whose time evolution is governed by a hyperbolic Hamiltonian. 

Radial conformal symmetries of Minkowski space-time have been studied and classified in \cite{RCM}. These conformal Killing vectors, according to the sign of their norm, divide Minkowski space-time into subregions with null boundaries.  It has been recently appreciated \cite{DeLorenzo:2017tgx,DeLorenzo:2018ghq} that the structure of these regions is analogous to the space-time associated to the inner and outer horizons of a Reissner-Nordstrom black hole. In fact one can associate to the null boundaries generated by radial conformal Killing vectors thermodynamical  relations analogous to the one satisfied by black holes. Causal diamonds can be seen as one type of these covariant subspaces. The conformal KIlling vectors generating transformations which map diamonds into themselves are a special class of radial conformal Killing vectors given by a combination of the generators of time translations and special conformal transformations. The integral curves of these diamond Killing vectors are worldlines of observers moving with constant acceleration and indeed their worldlines can be obtained from a conformal mapping of the trajectories followed by accelerated observers in the Rindler wedge \cite{Martinetti:2002sz,Martinetti:2008ja}. As a consequence one naturally expects these diamond observers to experience phenomena like the Unruh effect whereby the vacuum state for inertial observers appears to accelerated observers as a thermal state. Arguments in support of the existence of a diamond temperature associated to these observers were given for the first time in \cite{Martinetti:2002sz} based on a conjecture known as {\it thermal time hypothesis}. More recently, in \cite{Su:2016wop}, thermal effects associated to the restriction of the space-time domain to a causal diamond where studied introducing a set of Rindler-like diamond coordinates and analyzing the response function of a Unruh-DeWitt detector with Hamiltonian appropriately rescaled in order to be a constant of motion under the diamond time evolution. In this note we illustrate how the connection between diamond time evolution and hyperbolic time evolution in conformal quantum mechanics provides further evidence for the existence of a diamond temperature. Our analysis also shows that, within the framework of conformal quantum mechanics, thermal effects associated to a notion of time evolution which effectively restricts the time domain to a finite region can be derived starting from representations of the $SL(2,\mathbb{R})$ group, thus providing an example of quantum system in which a phenomenon in essence analogous to the Unruh effect can be rooted on group theoretic grounds. 

In the next Section we exhibit the connection between radial conformal Killing symmetries in Minkowski space-time and time evolution in conformal quantum mechanics. In particular we show that conformal Killing vectors which map a causal diamond into itself are related to the generator of {\it hyperbolic} time evolution in conformal quantum mechanics. In Section 3 we notice that two-point functions of conformal quantum mechanics, seen as a $CFT_1$, which express correlations of quantum states whose time evolution is governed by a hyperbolic Hamiltonian have a thermal character. The correspondence between this time evolution and diamond time is then used to connect such two-point functions, for a particular value of the conformal weight, with the one for a massless field along the diamond static trajectory and thus connecting the diamond temperature with the hyperbolic temperature in $CFT_1$. The final Section 4 is devoted to a brief summary and outlook.

\section{Radial conformal symmetry as time evolution}

Let us consider the line element of Minkowski space-time in spherical coordinates
\be\label{minksph}
d s^2 = -d t^2 + d r^2 + r^2 d \Omega^2\,,
\ee
where $d \Omega^2 = d \theta^2 + \sin^2 \theta\,  d\phi^2$. 
The most general {\it radial conformal} Killing vector  \cite{RCM}, i.e. a radial vector field $\xi$ such that the Lie derivative $\mathcal{L}_{\xi}$ of the Minkowski metric $\eta_{\mu\nu}$ is proportional through a constant factor to the metric itself $\mathcal{L}_{\xi} \eta_{\mu\nu} \propto \eta_{\mu\nu}$,  is given by
\be\label{xi2}
\xi = \left(a(t^2+r^2)+b t +c\right)\, \partial_t + r (2 a t + b)\, \partial_r\,,
\ee
where $a$, $b$, and $c$ are arbitrary {\it real} constants. Such conformal Killing vector can be written as a linear combination of generators of time translations 
\be\label{p0}
P_0 = \partial_t\,,
\ee 
dilations\footnote{We denote the generator of dilations with $D_0$ simply for notational convenience, the subscript $0$, unlike for $P_0$ and $K_0$, does not refer in this case to a time direction} 
\be\label{d0}
D_0 = r\, \partial_r + t\, \partial_t\,,
\ee 
and special conformal transformations along the time direction 
\be\label{k0}
K_0 = 2 t r\, \partial_r + (t^2+r^2)\, \partial_t = 2 t D_0 + (r^2-t^2) P_0\,, 
\ee
as
\be
\xi = a K_0 + b D_0 + c P_0\,.
\ee
Notice that these generators close the $\mathfrak{sl}(2,\mathbb{R})$ algebra 
\be\label{sl1}
[P_0,D_0]= P_0\,,\qquad [K_0,D_0]= - K_0\,,\qquad [P_0,K_0]= 2 D_0\,.
\ee
The starting observation of this note is that time evolution in conformal quantum mechanics \cite{deAlfaro:1976vlx} is determined by a Hamiltonian which in its most general form is given in the literature as
\be
G = u H + v D + w K
\ee
where $H$, $D$ and $K$ again belong to the $\mathfrak{sl}(2,\mathbb{R})$ algebra
\be\label{sl2}
[H,D]=i H\,,\qquad [K,D]= - i K\,,\qquad [H,K] = 2 i D\,.
\ee
More precisely, the generators $H$, $D$ and $K$ are the ``quantum mechanical" counterparts of $P_0$, $D_0$ and $K_0$ in the sense that $H= i P_0\,,\,\,K= i K_0   \,,\,\,D = i D_0$, 
and thus $G=i \xi$ if we make the identification $u=c$, $v=b$ and $w=a$. Such generators are strictly related to the isometries of $AdS_2$ and indeed $P_0$, $D_0$ and $K_0$, as given in \eqref{p0},  \eqref{d0} and \eqref{k0}, can be identified with the Killing vectors of $AdS_2$ in Poincar\'e coordinates \cite{Ho:2004qp,Jarvela:2015zra}.\\ 

There are three types of Hamiltonians which can not be connected to another via the adjoint action of $SL(2,\mathbb{R})$ \cite{deAlfaro:1976vlx,Ho:2004qp,Tada:2017wul} and thus belong to different conjugacy classes. Such classification can be obtained by noting that the invariance of the $\mathfrak{sl}(2,\mathbb{R})$ Casimir
\be
\mathcal{C}= \frac{1}{2}\left(K H + H K \right) - D^2
\ee
is reflected in the invariance of the determinant of the matrix 
\be
\begin{pmatrix}
   b & 2 c \\
   2 a & b \\
  \end{pmatrix}\,,
\ee
and the various Hamiltonians in conformal quantum mechanics can be classified according to the sign of $\Delta = b^2 - 4 a c\,$: 

\begin{itemize}
\item[-] $\Delta < 0$: these are generators of {\it elliptic} transformations. The generator of rotations
\be
R = \frac{1}{2} \left( \alpha H +  \frac{K}{\alpha} \right)\,,
\ee
falls into this class. Notice the introduction of the constant $\alpha$ with dimensions of length, needed to render $R$ dimensionless which will be crucial for what follows. The spectrum of such Hamiltonian is discrete and a basis of eigenstates of $R$ was introduced in \cite{deAlfaro:1976vlx} and used in \cite{Chamon:2011xk,Jackiw:2012ur} to construct the two-point functions of the theory which we will use below. One of the main upshots of this letter will be to give a geometrical interpretation of the constant $\alpha$.

\item[-] $\Delta = 0$: these operators generate {\it parabolic} transformations or, if we look at the correspondence between $SL(2,\mathbb{R})$ and the three-dimensional Lorentz group $SO(2,1)$, {\it null rotations}. Representatives of this class are $H$ and $K$.

\item[-] $\Delta > 0$: these are generators of {\it hyperbolic} transformation or boosts from the Lorentz group perspective. The generator of dilations $D$ and the combination
\be
S = \frac{1}{2} \left(\alpha H -  \frac{K}{\alpha} \right)\,,
\ee
which will be central for what follows, fall into this class.
\end{itemize}

The corresponding classification from the point of view of radial conformal Killing vectors in Minkowski space was given in \cite{RCM}. Such classification deals with the sign of the Minkowski norm of the conformal Killing vector which turns out to be determined by $\Delta$. Indeed it can be shown that \cite{RCM} (see also \cite{DeLorenzo:2017tgx, DeLorenzo:2018ghq})

\begin{itemize}
\item[-] For $\Delta < 0$ the radial conformal Killing vector is {\it everywhere timelike}.

\item[-] For $\Delta = 0$ the radial conformal Killing vector is also everywhere timelike except that for the light cone emanating from $t=-(b/2a), r=0$.

\item[-] For $\Delta > 0$ it is null on the light cones at the points $t=t_{\pm}, r=0$ where $t_{\pm} = \frac{- b \pm \sqrt{\Delta}}{2a}$, time-like inside either lightcone or outside {\it both} light cones and space-like everywhere else.

\end{itemize}
Notice that setting $b=0$ just makes everything symmetric about the origin $t=0, r=0$.\\

Now let us consider the time evolution determined by the three representative Hamiltonians $H$, $R$ and $S$. These cases were already considered in \cite{deAlfaro:1976vlx}. If we write 
\be
H = i \partial_t 
\ee
one can define new time variables $T$ and $\tau$ such that 
\be
R = i \partial_T\,, \qquad S= i \partial_\tau\,.
\ee
These three variables are related in conformal quantum mechanics through \cite{deAlfaro:1976vlx}
\be\label{tvarch}
t = \alpha \tan(T/2) = \alpha \tanh(\tau/2)\,.
\ee
An analogous association can be made with coordinates in Minkowski space. To see this let us notice that the generator $S =  i \partial_\tau$ of translations  in the time variable $\tau$ in conformal quantum mechanics can be written as a differential operator in Minkowski space-time in spherical coordinates as
\be\label{skill}
S = \frac{1}{2 \alpha} \left(\alpha^2 H - K \right) = \frac{i}{2 \alpha} \left((\alpha^2-t^2-r^2) \partial_t - 2 t r \partial_r \right)\,.
\ee
This generator corresponds to the special class of radial conformal Killing vectors which map a causal diamond, i.e. the region $|t|+|r|<\alpha$, into itself \cite{DeLorenzo:2017tgx, DeLorenzo:2018ghq,Visser:2019muv}
\be
\zeta = \frac{1}{2 \alpha} \left((\alpha^2-t^2-r^2) \partial_t - 2 t r \partial_r \right)\,.
\ee
Through this correspondence we see that the length scale $\alpha$ introduced in conformal quantum mechanics is related to the size of a causal diamond in Minkowski space, indeed $2 \alpha$ is the length of the vertical axis of the diamond, or the {\it finte lifetime} of a static observer sitting at the origin.\\ 

In spherical coordinates the Hamiltonian $H = i \partial_t$ is simply the generator of translations in the inertial time coordinate $t$. A set of coordinates in Minkowski space ``adapted" to the radial Killing vector $\zeta= -i S$, i.e. such that 
$i \zeta = S= i \partial_\tau$, is given by
\bea\label{diac}
t&=& \alpha\frac{\sinh \tau}{\cosh x + \cosh \tau}\\
r&=& \alpha\frac{\sinh x}{\cosh x + \cosh \tau}\,.
\eea
The Minkowski line element in this {\it diamond coordinates} reads
\be
d s^2 =\frac{\alpha^2}{(\cosh \tau + \cosh x)^2} \left(-d \tau^2 + d x^2 + \sinh^2 x\, d \Omega^2\right)\,,
\ee
which is evidently conformally related to the metric \eqref{minksph}. \\

Integral curves of a generic radial conformal Killing vector were obtained in \cite{RCM}. In the special case of the conformal Killing vector \eqref{skill} (i.e. when $a = - \frac{1}{2 \alpha}$, $b=0$ and $c=\frac{\alpha}{2}$ in \eqref{xi2}) they are given by the one-parameter family of hyperbolae in the $r-t$ plane
\be\label{wldo}
t^2 - (r- \alpha\, \omega)^2 = \alpha^2 (1-\omega^2)
\ee
parametrized by $\omega\neq 0$.  Each one of these curves\footnote{The curves \eqref{wldo} correspond to the ones found in \cite{Martinetti:2002sz} by mapping the worldlines of Rindler observers, i.e. orbits of boosts, to the diamond via a special conformal transformation.} can be seen as the worldline of an accelerated observer with finite lifetime confined to the causal diamond of size $2 \alpha$. From the four-velocity of observers static with respect to diamond time, i.e. observers whose worldlines are given by the curves \eqref{wldo}, one can derive the acceleration 
\be
a(x)=\frac{\sinh x}{\alpha} = \frac{1}{\alpha\, \sqrt{\omega^2-1}}\,,
\ee
which is {\it constant} along each integral curve and is related to the parameter $\omega = \frac{1}{\tanh x}$. Thus a translation in the diamond variable $x$ maps worldlines of observers with different acceleration into another. Diamond observers passing through the origin $x=0$ have vanishing acceleration and follow the same worldline of a static inertial observer sitting at the origin. From \eqref{diac} we see that inertial time and diamond time at $x=0$ are related by 
\be\label{indiam}
t = \alpha\, \frac{\sinh \tau}{1 + \cosh \tau} = \alpha \tanh(\tau/2)
\ee 
i.e. the same relationship between $H$-time and $S$-time variable in conformal quantum mechanics!

\section{Diamond temperature on the real line}

In Minkowski space-time the vacuum state associated to the free Hamiltonian of inertial observers is perceived as a thermal state populated of excitations associated to the ``boost" Hamiltonian generating time evolution for uniformly accelerating observers. This is the celebrated Unruh effect \cite{Unruh:1976db}. In \cite{Arzano:2018oby} it was shown how, also in the simplest case of a quantum system with affine symmetry living in {\it one dimension}, the freedom in the choice of translation generators leads to an analogous effect. In this case one considers particles as irreducible representations of the group of affine transformations of the real line, also known as the $ax+b$ group. This is comprised of translations and dilations. Particles can carry an ``energy" associated to the generators of both transformations. It turns out that the vacuum state with no excitations with respect to the translation Hamiltonian contains a thermal distribution of excitations associated to the dilation Hamiltonian.\\

A natural question that arises is whether a similar effect can be identified in a one dimensional quantum system with a larger group of symmetries. This is the case of conformal quantum mechanics \cite{deAlfaro:1976vlx}. Such model can be seen as a $0+1$-dimensional field theory like the one considered in \cite{Arzano:2018oby} in which, besides translations and dilations, one also considers special conformal transformations i.e. the group generated by the algebra \eqref{sl2}. The standard construction \cite{deAlfaro:1976vlx} of this field theory starts from the introduction of a set of eigenstates of the generator of roations $R$, $\ket{n}$, carrying an irreducible representation of $SL(2, \mathbb{R})$
\begin{subequations}
\begin{gather}
R \ket{n} = r_n\, \ket{n}\label{rjAds6-a}\\
r_n = r_0 + n,\ \  r_0 > 0, \ \ n = 0,1 \ldots \nonumber\\
L_\pm \ket{n} = \sqrt{r_n \, (r_n \pm 1) - r_0 \, (r_0 -1)} \, \ket{n \pm 1}\,,
\end{gather}
\end{subequations}
with inner product $\braket{n}{n^\prime} = \delta_{n\, n^\prime}$, where 
\be
L_\pm  \equiv  \frac{1}{2}\, \left(\frac{K}{\alpha} - \alpha\, H \right)\, \pm i\, D,  
\ee
and $|n=0\rangle$ is the ``R-vacuum". The constant $r_0$ is related to the eigenvalue of the Casimir operator of $\mathfrak{sl}(2, \mathbb{R})$
\be\label{casr0}
\mathcal{C}= \frac{1}{2}\left(K H + H K \right) - D^2 = r_0 \, (r_0 -1)\,.
\ee

We now look for states $\ket{t}$ labelled by a time variable $t$ on which $H$ acts as a generator of translations \cite{deAlfaro:1976vlx}
\be
H\ket{t} = - i\, \frac{d}{dt} \ket{t}\,. 
\ee
The action of the other generators on these states in terms of time derivatives will be given by
\bea
D \ket{t} &=& - i\, \left(t\, \frac{d}{d t} + r_0 \right)  \ket{t},\\
K \ket{t} &=& -i \left(t^2\,  \frac{d}{d t} + 2\, r_0\, t \right) \ket{t}\,
\eea
so that they close the algebra \eqref{sl2} and are related through the Casimir \eqref{casr0}. Notice how translations generated by $H$ can cover the entire time line $\mathbb{R}$, in analogy with the free Hamiltonian of inertial observers who have access to the whole Minkowski space-time.

The $\ket{t}$ states can be characterized by looking at their overlap with $\ket{n}$ states. Using the fact that the latter are diagonal under the action of the generator $R$ one can write the following differential equation
\be
\langle t|R|n\rangle = r_n\langle t |n\rangle = \frac{i}{2}\left[\left(\alpha+\frac{t^2}{\alpha}\right)\frac{d}{dt}+2r_0\frac{t}{\alpha} \right]\langle t |n\rangle
\ee
which can be solved to give \cite{deAlfaro:1976vlx}
\be\label{nts}
\langle t |n\rangle = (-1)^n\left[\frac{\Gamma(2r_0+n)}{n!}\right]^{\frac{1}{2}}\left(\frac{\alpha-it}{\alpha+it}\right)^{r_n}
\left(1+\frac{ t^2}{\alpha^2}\right)^{-r_0}\,.
\ee
Alternatively \cite{Jackiw:2012ur}, the kets $\ket{t}$ can be obtained directly from the $R$-vacuum through a ``complex" time translation 
\begin{equation}\label{tket}
\ket{t} = e^{i H t}\, \ket{t=0} = \frac{\Gamma^{1/2}\, (2 r_0)}{2^{2 r_0}}\ e^{(\alpha + i t)\, H}\, \ket{n=0}\, .
\end{equation}
From \eqref{nts}, using the completeness of the states $\ket{n}$, one can derive the inner product of $\ket{t}$ states
\be\label{innt}
 \braket{t_1}{t_2} = \frac{\Gamma\, (2 r_0)\, \alpha^{2 r_0}}{[2 i\, (t_1 - t_2)]^{2 r_0}}\,,
\ee
which are thus {\it not} orthonormal. In \cite{Chamon:2011xk} the authors propose to identify the inner product \eqref{innt} with the two-point functions of a $CFT_1$ 
\be
G_2 (t_1, t_2) \equiv \braket{t_1}{t_2} = \frac{\Gamma\, (2 r_0)\, \alpha^{2 r_0}}{[2 i\, (t_1 - t_2)]^{2 r_0}}\,,
\ee
with conformal weight $r_0$.
For $r_0 = 1$ such two-point function is identical, modulo a constant factor, to the two-point function of a free massless scalar field in Minkowski space-time in the inertial vacuum, evaluated along the trajectory of a static inertial observer sitting at the origin. In other words two-point correlators for a massless field along the worldline of static observers in Minkowski space-time are in correspondence with two-point functions of conformal quantum mechanics for the states $\ket{t}$.\\

There is a subtlety concerning the states $\ket{t}$, discussed in detail in \cite{Chamon:2011xk,Jackiw:2012ur}. As mentioned above, such states can be obtained by acting on the $R$-vacuum with the operator
\be
\mathcal{O}(t)= \frac{\Gamma^{1/2}\, (2 r_0)}{2^{2 r_0}}\ e^{(\alpha + i t)\, H}\,.
\ee
As it was already noticed in \cite{deAlfaro:1976vlx}, the $R$-vacuum is not invariant under the action of the full conformal group $SL(2,\mathbb{R})$. At the same time the operator $\mathcal{O}(t)$ is not a conformal primary. However, as discussed in \cite{Chamon:2011xk}, these two ``flaws" compensate each other when $\mathcal{O}(t)$ acts on the $R$ vacuum and creates states $\ket{t}$ which are conformally covariant. From a state-operator-correspondence point of view the operator $\mathcal{O}(t)$ when acting on the $R$-vacuum behaves as a conformal primary of weight $r_0$. This observation has a crucial upshot for our discussion since it tells us that under a change of time variable from $t$ to a new variable $t'$ the quantum states transform as
\be
  \ket{t' }= \left( \frac{dt}{d t'}\right)^{r_0} \ket{t=t(t')} \,.
\ee
Let us now look at the action of the generator S on the kets $\ket{t}$
\begin{equation}
S \ket{t} = \frac{1}{2}\ \left(\alpha H - \frac{K}{\alpha}\right)\ \ket{t} = -i \left(\frac{1}{2} \ {[t^2/\alpha - \alpha]}\ \frac{d}{d t} + \frac{r_0 t}{\alpha}\right)\ \ket{t}\,.
\end{equation}
As seen in Section 2, $S$ generates translations in the time variable $\tau$ related to $t$ through \eqref{tvarch} 
\be
t = \alpha \tanh(\tau/2)\,.
\ee
Accordingly we will have new states $\ket{\tau}$ associated to this time variable
\begin{equation}\label{ketau}
\ket{\tau} = \left(\frac{\alpha}{2}\right)^{r_0}(\cosh \tau/2)^{-2 r_0} \ket{t = \alpha \tanh (\tau/2)},
\end{equation}
such that \cite{Jackiw:2012ur}
\begin{equation}
S \ket{\tau} = - i\, \frac{d}{d \tau}\ \ket{\tau}\,.
\end{equation}
The two-point function associated to the new kets is then given by 
\begin{eqnarray}\label{2ptau}
G_2 \, (\tau_1, \tau_2) = \braket{\tau_1}{\tau_2} = \frac{\Gamma (2 r_0)\, \alpha^{2r_0}}{\left[ 4 i \sinh \left(\frac{\tau_1 - \tau_2}{2}\right)\right]^{2r_0}}\ .
\end{eqnarray}
This two-point function is periodic in imaginary time and it is associated to a thermal state at temperature $T=\frac{1}{2 \pi}$. Such thermal character has been linked in the literature to the temperature perceived by a static patch observer in de Sitter space \cite{Anninos:2011af,Nakayama:2011qh}.\\

Here we show that there is a direct connection between the thermal two-point function of conformal quantum mechanics \eqref{2ptau} and the one associated to a static diamond observer. Let us first note that the time variable $\tau$ associated to the $S$ generator, as customarily defined in conformal quantum mechanics, i.e. as in equation  \eqref{tvarch}, is {\it dimensionless}. The right dimensions can be recovered by defining a new time variable multiplied by a factor containing the length scale $\alpha$. We can determine such factor by requiring the generator of diamond time translations to {\it match the generator of inertial time translations H} when the size of the diamond goes to infinity, i.e. defining a new generator $S'$ 
\be\label{diapr}
S' = \frac{2}{\alpha}\, S = H-\frac{K}{\alpha^2}\,. 
\ee
This choice can be realized by introducing new diamond coordinates $\tau' = \frac{\alpha}{2}\, \tau\,,\,\, x' = \frac{\alpha}{2}\, x$
so that $S'= i\partial_{\tau'}$ and with inertial time related to the new diamond time for worldlines passing through the origin via (see \eqref{indiam})
\be\label{ttaup}
t = \alpha \tanh(\tau'/\alpha)\,.
\ee
The choice of Hamiltonian \eqref{diapr} is justified since $S' $ generates time translations for the static diamond observer and when the observer's lifetime goes to infinity, $\alpha\rightarrow \infty$, her notion of time coincides with that of a static inertial observer $t = \tau'$ (as it is evident from \eqref{ttaup}). Similarly for what happens with translations associated to the dilation generator in the case of the affine group \cite{Arzano:2018oby}, here the time translations associated to the generators $S$ and $S'$ only cover a portion $t\in (-\alpha,\alpha)$ of the time line when $\tau$ and $\tau'$ range from $-\infty$ to $\infty$. In analogy with the change of variable from $t$ to $\tau$ we can now define new kets $\ket{\tau'}$ whose two-point function for $r_0=1$ is given by
\begin{eqnarray}\label{2ptaupri}
G_2 \, (\tau^\prime_1, \tau^\prime_2) = -\frac{\alpha^2}{16 \sinh^2 \left(\frac{\tau^\prime_1-\tau^\prime_2}{\alpha}\right)}\ .
\end{eqnarray}
This is a thermal two-point function at temperature
\be\label{dtemp}
T_{S'} =  \frac{1}{\pi \alpha}
\ee
which is exactly the diamond temperature for a finite lifetime observer found in \cite{Martinetti:2002sz} using the {\it thermal time hypothesis}\footnote{It should be noted however that the diamond temperature obtained in \cite{Martinetti:2002sz} is time dependent and that the temperature \eqref{dtemp} is the one perceived by the observer in the middle of her lifetime i.e. at $\tau'=0$.}. The two-point function \eqref{2ptaupri} coincides (modulo a $(2\alpha^2/\pi)^2$ factor) with the two-point function of a massless scalar field evaluated along the trajectory of a Rindler observer moving with constant acceleration $a=\frac{2}{\alpha}$ whose proper time is $\tau'$ \cite{Su:2016wop,Michel:2016qge,Dhumuntarao:2018way}. The coincidence of the two-point function of a massless scalar field written in Rindler-like diamond coordinates with such two point function was exploited in \cite{Su:2016wop} in order to derive the diamond temperature \eqref{dtemp} using a Unruh-DeWitt detector with a rescaled Hamiltonian. The effect of such rescaling is to multiply the two-point function in the Minkowski vacuum used to calculate the detector's response function \cite{Su:2016wop} by a non-trivial factor. This factor, from the point of view of the corresponding conformal quantum mechanics, results from the non-trivial transformation properties of the kets \eqref{ketau} when writing the two-point function in terms of the hyperbolic time variable $\tau$.

\section{Conclusions}
Causal diamonds play a fundamental role in the understanding of thermodynamic properties of space-time and gravity \cite{Gibbons:2007fd,deBoer:2016pqk,Jacobson:2018ahi}. According to the covariant entropy bound \cite{Bousso:1999xy} they also have holographic properties: one can associate a Bekenstein-Hawking entropy to the null boundary of the causal diamond. In this note we exhibited a deep connection between the radial symmetries of causal diamonds and time evolution in conformal quantum mechanics, a one-dimensional quantum field theory which appears to play an important role in a variety of contexts from cosmology \cite{BenAchour:2019ufa} to black hole thermodynamics \cite{Camblong:2004ye}, from holography \cite{Chamon:2011xk,Jarvela:2015zra} to the understanding of quantum anomalies \cite{Camblong:2003mz}. In particular we showed that the thermal behaviour of one-dimensional conformal two-point functions is directly connected with the two-point function along the worldline of a static diamond observer experiencing a temperature inversely proportional to her lifetime. It is left to future studies to further explore the significance of the effect we analyzed in this note for the disparate variety of contexts in which conformal quantum mechanics appears as an effective theory.

\section*{Acknowledgment}
I would like to thank F. Alessio and J. Kowalski-Glikman for useful discussion. I would also like to thank the Institute of Theoretical Physics at the University of Wroclaw for hospitality while this work was being finished.

\end{document}